\documentclass[useAMS,usenatbib]{mn2e}

\usepackage{amssymb}
\usepackage{aas_macros}
\usepackage{natbib}
\usepackage{times}
\usepackage[dvips]{graphicx}
\usepackage{color}
\usepackage{hyperref}

\usepackage[T1]{fontenc}
\usepackage{ae,aecompl}

 \usepackage{hyperref}

\providecommand{\adsurl}[1]{\href{#1}{ADS}}
  
\usepackage{abrev} 

\title[Simulated galaxies on bulge scaling relations]{Simulating disc galaxy bulges that are consistent with observed scaling relations}
\author[C. Christensen  et al. ]
{
\parbox[t]{\textwidth}{
 C.~R.~Christensen\thanks{E-mail:cchristensen@as.arizona.edu}$^{1}$, 
 A.~M.~Brooks$^{2}$,
 D.~B.~Fisher$^{3,4}$,
  F.~Governato$^{5}$,
 J.~McCleary$^6$,
 T.~R.~Quinn$^{5}$,
 S.~Shen$^7$,
 J.~Wadsley$^8$,
             }
\vspace*{6pt} \\
$^1$ Department of Astronomy, University of Arizona, 933 North Cherry Avenue, Rm. N204, Tucson, AZ 85721-0065, USA;\\
$^2$ Department of Physics \& Astronomy, Rutgers, The State University of New Jersey, 136 Frelinghuysen Rd, Piscataway, NJ 08854, USA;\\
$^3$ Centre for Astrophysics and Supercomputing, Swinburne University, Hawthorn VIC 3122, Australia;\\
$^4$ Department of Astronomy, University of Maryland, CSS Bldg., Rm. 1204, Stadium Dr., College Park, MD 20742-242, USA;\\
$^5$ Department of Astronomy, University of Washington, Box 351580, Seattle, WA 98195, USA;\\
$^6$ Department of Astronomy, New Mexico State University, P. O. Box 30001, MSC 4500 Las Cruces, NM 88003-8001USA;\\
$^7$ Department of Astronomy and Astrophysics, University of California, Santa Cruz, CA 95064, USA;\\
$^8$ Department. of Physics and Astronomy, McMaster University, Hamilton, ON L8S 4L8, Canada
\vspace*{-0.2cm}}

\begin{document}
\date{Accepted to MNRAS Letters}
\pagerange{\pageref{firstpage}--\pageref{lastpage}} \pubyear{2013}
\maketitle
\label{firstpage}

\begin{abstract}
We present a detailed comparison between the photometric properties of the bulges of two simulated galaxies and those of a uniform sample of observed galaxies.
This analysis shows that the simulated galaxies have bulges with realistic surface brightnesses for their sizes and magnitude.
These two field disc galaxies have rotational velocities $\sim$ 100 km/s and were integrated to a redshift of zero in a fully cosmological $\Lambda$ cold dark matter context as part of high-resolution smoothed particle hydrodynamic simulations.
We performed bulge-disc decompositions of the galaxies using artificial observations, in order to conduct a fair comparison to observations.
We also dynamically decomposed the galaxies and compared the star formation histories of the bulges to those of the entire galaxies.
These star formation histories showed that the bulges were primarily formed before {\em z} = 1 and during periods of rapid star formation.
Both galaxies have large amounts of early star formation, which is likely related to the relatively high bulge-to-disc ratios also measured for them.
Unlike almost all previous cosmological simulations, the realistically concentrated bulges of these galaxies do not lead to unphysically high rotational velocities, causing them to naturally lie along the observed Tully--Fisher relation.
\end{abstract}

\begin{keywords}
methods: numerical, galaxies: bulges, galaxies: formation, galaxies: spiral, galaxies: structure
\end{keywords}

\begin{table*}
\begin{center}
\begin{tabular}{l|ccccccccccccc}                                        
&				& Virial mass			& $V_f$	         & $M$	& $M$	&$B$/$T$		& $M_{\mathrm{Bulge}}$	& $r_e$ (pc)	& $\mu$ 	&  S\'ersic Index	& $M_{\mathrm{Disk}}$	& $r_h$ (pc)	\\
&				& ($\Msun$) 			& (km s$^{-1}$)		& ($B$)		&($V$)		 & ($H$)		& ($H$)			&($H$)	 		& ($H$)		 & ($H$) 			&  ($H$)		&  ($H$) 	\\ 
&				& 1					& 2			& 3		& 4		& 5			& 6			& 7			& 8		& 9				& 10			& 11			\\\hline \hline

&{\bf h603} 		& $3.4\times 10^{11}$	& 116		& $-19.45$	& $-19.92$ 	& 0.43		& $-20.86$		& 830		& 16.35	& 1.65			& $-21.17$		& 2700		\\ 
&{\bf h986} 		& $1.9\times 10^{11}$	& 107		& $-19.12$	& $-19.50$ 	& 0.53		& $-20.81$		& 2200		& 18.78	& 2.53			& $-20.68$		& 5400		\\ 
\end{tabular}
\end{center}
\caption[Properties of the a set of galaxies with different ISM models at z = 0]
{Properties of the main haloes at {\em z} = 0.
$V_f$ is the rotational velocity in the outer parts of the gas disc.
{\sc sunrise} was used to determine the {\em B} and {\em V}-band Vega magnitudes and the {\em H}-band photometric image.
A S\'ersic bulge and exponential disc were fit to the {\em H}-band profile.
The resulting bulge-to-total ratios ($B$/$T$), bulge magnitudes ($M_{\mathrm{Bulge}}$), half-light radii ($r_e$), surface brightness within $r_e$ ($\mu$), S\'ersic indies of the bulge, disc magnitudes ($M_{\mathrm{Disk}}$), and disc scale lengths ($r_h$) are listed in columns 5 - 11.}
\end{table*}

\section{Introduction}
Computational astronomers have made great strides in reproducing the observed disc structure of spiral galaxies in $\Lambda$ cold dark matter simulations, as studies of the Tully--Fisher relation \citep{Robertson04, Governato07, Stinson10,
Piontek11}, angular-momentum content \citep{Scannapieco09}, and disc sizes \citep{Brooks11} have shown.
Historically, though, the corresponding bulges of simulated galaxies were much too large and concentrated compared to observed bulges \citep{Bullock01, vandenBosch01b, Binney01, vdb02, donghia06, dn07,Stinson10, scannapieco11}. 
This difficulty in reproducing the bulges of disc galaxies represents the more-persistent aspect of the `angular-momentum catastrophe' \citep{Navarro94}.

In the past several years, only a handful of simulations have been able to combine the resolution necessary to resolve bulges in cosmological simulations with star formation and feedback models capable of producing galaxies with realistic mass bulges.
These simulations have variously succeeded by limiting the star formation efficiency \citep{Agertz10}, pre-heating the gas through early stellar feedback \citep{Stinson13}, or increasing the strength of supernova-driven (SN-driven) outflows, either by scaling a kinetic wind model \citep{Okamoto12, Vogelsberger13} or concentrating the energy through a high star formation threshold \citep{Guedes11, Brook11a}.
Similar strong outflows have been shown to reduce the amount of low-angular momentum material \citep{Pontzen11}, resulting in bulgeless or cored dwarf galaxies \citep{Oh11, Governato12,Teyssier13} and disc galaxies with realistic, rising rotation curves \citep{Christensen12a, AnglesAlcazar13}. 
Despite these improvements, no single method has been shown to reproduce the observed properties of bulges in a range of galaxies.
As such, the masses and concentration of bulges remain as firm limits on the formation of galaxies.
In particular, the observed scaling relations between bulge surface density, magnitude, and size encapsulate much of the observational constraints.

Given the importance of reproducing bulge properties, it is critical to analyse them in an observationally motivated fashion.
Here we undertake a careful photometric analysis of the bulges of two simulated galaxies and compare them to an observed sample analysed in an identical fashion.
These galaxies were previously shown to have flat rotation curves and a reduced mass of low-angular momentum material as the result of feedback-driven outflows \citep[][hereafter C12a]{Christensen12a}.
In this Letter, we show that they also have realistically concentrated bulges.

\section{Computational Methods and Analysis}\label{sec:method}
\subsection{Description of the Simulations}\label{sec:sims}

We analysed high-resolution cosmological simulations of two field disc galaxies with virial masses of 3.4 and 1.9 $\times 10^{11}$ $\Msun$.
These simulations are discussed in detail in C12a as the `$\Hmol$'-ISM model.
The simulations were integrated using the $N$-body smoothed particle hydrodynamic code, {\sc Gasoline} \citep{Wadsley04}, assuming a Wilkinson Microwave Anisotropy Probe 3 cosmology \citep{Spergel07}.
By using the `zoom-in' technique \citep{Katz92}, we were able to achieve force spline softening lengths of 170 pc and dark matter, stellar, and gas particle masses of 1.3$\times$10$^5$, 2.7$\times$10$^4$, and 8.0$\times$10$^3 \Msun$.
This resolution was sufficient for the creation of dense star forming regions and to resolve the bulges.

{\sc gasoline} tracks the non-equilibrium abundances of H and He, including $\Hmol$ \citep{Christensen12} and metal enrichment.
In addition to H and He cooling processes, {\sc gasoline} includes cooling from metal lines \citep{Shen10}.
A cosmic ultraviolet field \citep{Haardt96} is responsible for photoionization and heating, while a local Lyman Werner radiation field dissociates $\Hmol$.
Both H{\sc I} and $\Hmol$ are shielded by dust, and $\Hmol$ is self-shielded.
Star formation proceeds stochastically according to the dynamical time: $p = \frac{m_{gas}}{m_{star}}(1 - \mathrm{e}^{-c^* \Delta t /t_{form}})$, where $m_{gas}$ is the mass of the gas particle, $m_{star}$ is the initial mass of the potential star particle, $\Delta t$ is the star formation timestep, $t_{form}$ is the local dynamical time, and $c^*$ is a star-forming efficiency factor \citep{Stinson06}.
In order to incorporate the observed link between star formation and $\Hmol$ \citep[e.g.][]{Bigiel10}, the star formation efficiency of a gas particle depends directly on the molecular hydrogen fraction: $c^* = 0.1 X_{\Hmol}$ \citep{Christensen12}.
One of the results of limiting star formation to dense, molecular regions is that SN feedback energy is highly concentrated and, therefore, able to drive outflows \citep{Governato10}.
SN feedback is implemented according to the `blastwave' feedback scheme \citep{Stinson06}.
In this scheme, cooling is temporarily shut off for a time equal to the momentum-conserving phase of the SN blastwave to prevent artificial cooling of the gas particles.
Here, we assume that all of the canonical $10^{51}$ erg released per SN are transferred to the interstellar media.

\subsection[mockO]{Photometric Decompositions}
\label{sec:mocO}

Simulated observations of the galaxies were created using the Monte Carlo radiative transfer code, {\sc sunrise} \citep[v 3.6][]{Jonsson06}.
The magnitudes and photometric images of our galaxies in \citet{Johnson66} and Two Micron All Sky Survey \citep[2MASS][]{Cohen03} bands (Table 1) were calculated by convolving the generated spectral energy distribution with the filter transmission curves. 
We decomposed the {\sc sunrise}-generated photometric images of the galaxies into bulge and disc components using the same analysis discussed in detail in \citet{Fisher08}.
We opted to use the face-on 2MASS H-band images, as infrared data are less sensitive to the obscuring effects of dust and the face-on orientation provides the most accurate fit.  
We determined the surface brightness profiles through ellipse fitting using the routine of \citet{Bender87}.
During this ellipse fitting, the radial sizes of ellipses were optimized to maintain a roughly constant signal-to-noise ratio across the profile. 
For every galaxy, we also fitted isophotal ellipses to the image with the software of \citet{Lauer85}, which is a Fourier based ellipse fitting method.
We then averaged together the isophotes from both the \citet{Bender87} and \citet{Lauer85} routines.

The bulge-disc decompositions were determined by fitting two components, a S\'ersic bulge plus outer exponential disc, to the major axis surface brightness profile. 
The inner radius cut for the fit was chosen to be 200~pc (approximately the softening length), and the outer radius cut was set to the location where the profile broke from a smoothly varying exponential.
The values of the bulge and disc magnitudes, the half-light radius of the bulge ($r_e$), the mean surface brightness within $r_e$($\mu$),  the disc scale length ($r_h$), the S\'ersic index, and the $B$/$T$ ratio are listed in Table 1.
Neither galaxy shows evidence of a bar at a redshift of zero.

While photometric decomposition is key to comparing simulations to observations, kinematic decomposition is necessary to separate the individual star particles into the bulge, disc, and halo components.
We defined the spheroidal component of the galaxies to be the stars whose specific angular momentum ($j_z$) was not a significant fraction of that of a circular orbit with the same binding energy, i.e., $j_z/j_c < 0.8$.  
We then subdivided the spheroidal component into the bulge and halo stars by making a radial cut where the spheroid mass profile changed to a shallower slope.
This process is outlined in greater detail in \citet{Governato09} (see also \citep{Abadi03,Scannapieco09}).

\section{Results}
\subsection{Bulge Photometric Properties}\label{sec:decomp}
\begin{figure}
\begin{center}
\includegraphics[width = 0.49\textwidth]{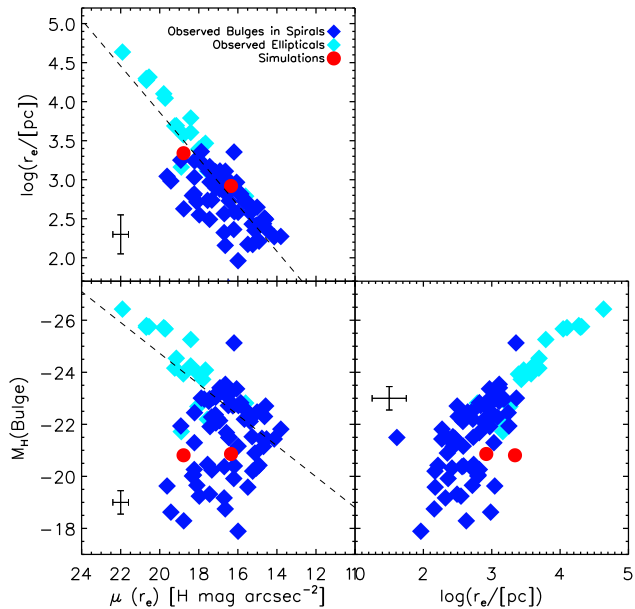}
\end{center}
\caption[Magnitude of the bulge versus effective radius]
{
Bulge scaling relations. 
The top panel shows the half light radii ($r_e$) of the bulges compared to the mean surface brightness within the half light radius [$\mu(r_e)$], also known as the \citet{Kormendy77} relation.
The bottom-left panel shows the magnitude of the bulge ($M_H$) versus $\mu(r_e)$ and the bottom-right panel shows $M_H$ versus $r_e$.
The simulated galaxies are represented by red circles.
Observational data are from \citet{Fisher10} and \citet{Fisher13}, and marked by dark blue (spiral galaxies) and light blue (elliptical galaxies) diamonds.
The black error bars are for the observed data and the dashed lines are for the observed scaling relations of elliptical galaxies.
The simulated galaxies lie within the observed relations, aside from one boarder-line point in the bottom-left panel, indicating that their bulges have appropriate concentrations.
}
\label{fig:BD}
\end{figure}

Elliptical galaxies follow robust scaling relations between surface brightness, luminosity, and half-light radii.
The bulges of disc galaxies, however, are more varied.
While classical bulges follow the elliptical galaxy scaling relations, pseudo-bulges tend to have lower surface brightnesses for the same size or magnitude \citep{Gadotti09,Fisher10}.
In contrast to both classical and pseudo-bulges, simulated bulges have historically been too bright and concentrated, leading them to have {\em higher} surface brightnesses than predicted by the observed scaling relations.

Here, we compare the photometric properties of the simulated bulges to those of an observed sample of disc and elliptical galaxies, and the associated scaling relations \citep{Fisher10, Fisher13}.
This sample includes disc galaxies from Hubble types of Sa-Sd and, therefore, covers the range of bulge properties. 
It also covers a similar range in $B - V$ as large surveys of galaxies, such as \citet{DeVaucouleurs91}, and preferentially samples bright and large disc galaxies in which bulges are more commonly found.
Critically, the photometric decompositions of the simulated galaxies followed exactly the same process as the observed galaxies.
Fig.~\ref{fig:BD} shows the bulge $H$-band surface brightnesses, magnitudes, and half-light radii of the simulated and observed galaxies.
The simulated galaxies lie among the observed galaxies when examining both size versus surface brightness and magnitude versus surface brightness, indicating that the bulges of the simulated galaxies are not overly concentrated.
One of the galaxies, h986, lies slightly outside the observed galaxies in the magnitude vs. size plot, although within the error.
The relatively large size of its bulge indicates that, if anything, it is not concentrated enough and may be evidence of over-correction on the part of the simulations.

The S\'ersic indices of the simulated galaxies (1.65 and 2.53) are consistent with those determined for the observed sample of disc galaxies, which had a median value of 1.86 and a standard deviation of 1.0.
In addition to examining the bulge properties, we also determined the disc scale-lengths and magnitudes.
The disc scale-lengths for similar magnitude discs in the observed sample range between $800$ and $3900$pc.
The disc scale-length of the simulated galaxy, h603, is $2700$ pc and consistent with these disc scale lengths.
The disc scale-length of h986 is highly uncertain and can vary by a factor of several.
The length listed in Table 1 ($5,400$ pc) is larger than that of galaxies with similar disc magnitudes.
However, when $B$-band images of it were decomposed, the disc was found to be consistent with the observed size vs. magnitude relation (Brooks et al. {\em in preparation}).

Both simulated galaxies have large $B$/$T$ values in the $H$ band: 0.43 and 0.53.
Because of the lack of recent star formation in the bulge, these values are lower in bluer bands. 
For instance, decompositions of $B$-band images yield $B$/$T$ values of 0.28 and 0.34.
While not entirely unheard of for galaxies of this mass, these ratio are at the far upper extreme.
For instance, in the comparison observational sample, most galaxies with magnitudes within 0.3 mag of the simulated galaxies had $B$/$T$ values between 0.05 and 0.1, with only one outlier having a $B$/$T$ equal to 0.4.
This discrepancy indicates that while the {\em shape} of the bulges is correct, other processes may be required to reduce the bulge {\em mass}.

\subsection{Star formation histories}\label{sec:sfh}

\begin{figure}
\begin{center}
\includegraphics[width = 0.5\textwidth]{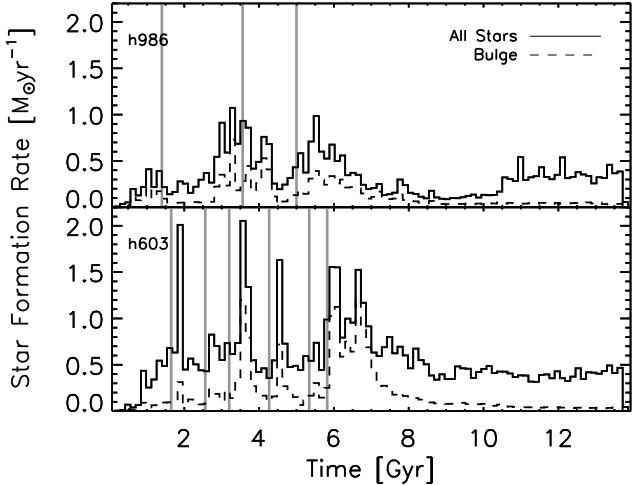}
\end{center}
\caption[Star Formation History]
{
SFHs of the bulges of the galaxies (dashed lines), compared to those of the entire galaxies (solid lines).
The grey bars indicate the beginnings of mergers with mass ratios of 10-to-1 or higher.
Bulge star formation happens preferentially during the first half of galaxy evolution and during periods of peak star formation. 
In both galaxies, it declines following the last merger.
}
\label{fig:sfh}
\end{figure}

Bulges have older average stellar ages than discs, implying that a large fraction of their stellar mass was formed comparatively early \citep{Allen06, MacArthur09}. 
Furthermore, delayed star formation results in more gas-rich mergers and smaller bulges \citep{Hopkins09a}.
The bulge properties should, therefore, be related to the star formation histories (SFH) of the galaxies and their bulges (Fig.~\ref{fig:sfh}).
We used kinematic decomposition to select individual bulge star particles, which resulted in a slightly different component selection than photometric decomposition.
For instance, the $B$/$T$ ratios produced by the kinematic decompositions were 0.57 and 0.56, as opposed to 0.43 and 0.53 (see \citet{Scannapieco10}).

During periods of peak star formation (generally following the onset of mergers, as indicated by grey bars in the plot), bulge stars were predominantly formed. 
In C12a, we found that during such peak star formation times, SN-driven outflows were very effective at driving out low-angular-momentum gas, which would have limited bulge growth.
Additionally, almost all bulge star formation happened within the first 7 Gyr for both galaxies.
While both galaxies lie along the observed redshift-zero halo--stellar mass relation \citep{Munshi12}, abundance matching from  \citet{Moster13} indicates that these simulations may have excessive high-$z$ star formation.
For instance, \citet{Moster13} predict that for galaxies of this mass, half the stellar mass will have been formed at redshift 0.25, whereas for these simulations, half the stellar mass was formed by redshifts 1.15 and 1.00.
Reducing the amount of high-redshift star formation would likely bring the B/T ratios closer to the observed values.

Examining the bulge SFHs raises the question of whether these galaxies have classical bulges or pseudo-bulges.
The SFHs of both simulated galaxies show at least some evidence of the merger-driven growth associated with classical bulges \citep{Kormendy13}.
The position of both simulated galaxies along the scaling relations is typical of classical bulges but still consistent with pseudo-bulges.
These two factors combined with a S\'ersic index of 2.53 indicate that h986 is most likely a classical bulge.
The lower S\'ersic index of h603 (1.65), however, is more consistent with a pseudo-bulge.
When considered in light of its heightened star formation following mergers and its position along the scaling relations, h603 could be a composite pseudo-classical bulge \citep{Fisher10}.

\subsection{Tully--Fisher Relation}

\begin{figure}
\begin{center}
\includegraphics[width=0.5\textwidth]{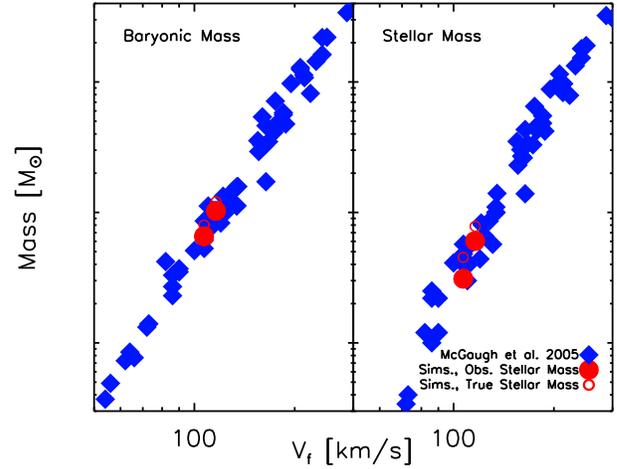}
\end{center}
\caption[Baryonic Tully--Fisher relation for simulated galaxies]
{ 
Baryonic and stellar Tully--Fisher relations.
Simulation stellar masses were determined both from mock-photometry (filled red circles) and directly from the simulations (empty red circles).
Both simulated galaxies lie along observed data from \citet{McGaugh05} (filled blue diamonds).
}
\label{fig:tf}
\end{figure}

In order to ensure that reducing the central concentration of the bulge through feedback did not disrupt the discs, we compared the simulated galaxies to the observed stellar and baryonic Tully-Fisher relations from \citet{McGaugh05} (Fig.~\ref{fig:tf}).
As in \citet{McGaugh05}, stellar masses were determined from the {\sc sunrise}-generated $B$-band magnitudes and $B - V$ colours of the galaxies, and gas masses were calculated by totalling the H{\sc i} mass within the disc and multiplying by 1.4 to account of He{\sc i}.
We also include data points representing the actual stellar mass; the difference between the actual and observationally determined stellar masses was previously discussed in \citet{Munshi12}.
$V_f$ was defined to be the average rotational velocities of gas in the outer portion of the gas disc (the radii that contained 80--90\% of the disc gas).
As these galaxies have flat rotation curves (C12a), $V_f$ was largely insensitive to the measurement radii.
Both simulated galaxies lie along the baryonic and stellar Tully--Fisher relations, indicating that, in addition to realistically concentrated bulges, they have appropriate disc stellar and gas masses.
The match to the observed relations is also a strong indication that the simulated galaxies had the correct distribution of matter -- too high a concentration of baryons would have resulted in a peaked rotation curve and too large of $V_f$ for its stellar and baryonic masses.

\section{Conclusions}\label{sec:res5}
In this Letter, we have taken an observationally motivated approach to analyzing the bulges of two simulated galaxies, previously shown to have rising rotation curves (C12a).
We photometrically decomposed H-band images into bulge and disc components and compared the bulge properties to an observational sample of disc and elliptical galaxies.
We determined that these simulations had bulges of the appropriate surface brightness for their magnitude and size.
This success indicates that the centres of the simulated galaxies have appropriate stellar distributions and that they are not overly concentrated.
Their relatively high bulge-to-total ratios, however, remain a concern and demonstrate a need for a further reduction in the central stellar mass.
In particular, reducing the amount of early star formation could both lower the bulge mass and result in more realistic SFHs.

We compared the SFHs of the kinematically selected bulges to those of the entire galaxies.
We found that the bulges formed primarily during the first half of the galaxies' lifetimes and that bulge star formation occurred during periods of peak star formation, possibly driven by mergers.
Of the two galaxies, the SFHs and photometric properties of one were most consistent with the properties of classical bulges, while the other shared characteristics with both classical and pseudo-bulges.
Finally, we verified the global structure of the simulated galaxies by comparing them to the observed baryonic and stellar Tully--Fisher relations.

\section*{Acknowledgements}
The authors would like to thank the anonymous referee for the suggestions.
These simulations were run at  NASA AMES and Texas Supercomputing Center. 
CC acknowledges support from NSF grants AST-0908499 and AST-1009452.
FG acknowledges support from HST GO-1125 and NSF AST-0908499.
TQ acknowledges support from NSF grant  AST-0908499.
We made use of pynbody (\url{https://github.com/pynbody/pynbody}) in our analysis.

\small
\bibliography{ObsBulges.bib}

\label{lastpage}

\end{document}